\documentstyle[12pt,epsfig]{article}
\begin{document}
\begin{titlepage}

\large
\centerline {\bf Beauty97 Conference Summary}
\normalsize
 
\vskip 2.0cm
\centerline {Samim~Erhan\footnote{Email: samim@uxucla.cern.ch}}
\vspace{3.0mm}
\centerline {\it University of California\footnote{Supported by U.S. National 
Science Foundation Grant PHY94--23142}, Los Angeles, CA  90095, U.S.A.}
\vskip 3.5cm
\centerline {\bf Abstract}
\vskip 1.0cm 

CP--violation is one of the least understood phenomena in our field. 
There are major experimental programs in all high energy laboratories 
around the world which will hopefully remedy this within the next decade.
The study of CP--violating effects in B meson decays will allow stringent 
tests of the Standard Model to be made and may point the way to New Physics.

The Beauty97 conference provided a forum for these experiments to discuss 
their physics potential and experimental challenges relating to these studies. 
This paper reviews the ongoing and future experimental B--physics projects.
I will summarize the status and future plans of these projects, as well as 
the highlights of the physics and key R\&D results presented at the 
conference. 
At the end, a critical comparison of the CP--violation B experiments will 
be given.

\vskip 1.0cm
\centerline {Nucl. Instrum. \& Methods, in press (1998).}
\vfill
\end{titlepage}

\clearpage\mbox{}\clearpage
\pagestyle{plain}
\setcounter{page}{1}

\def\NIM#1#2#3{Nucl. Instrum. \& Methods {\bf #1}\ (#2)\ #3}
\def\subNIM{submitted to Nucl. Instrum. \& Methods}
\def\NP#1#2#3{Nucl. Phys. {\bf #1}\ (#2)\ #3}
\def\PL#1#2#3{Phys. Lett. {\bf #1}\ (#2)\ #3}
\def\PR#1#2#3{Phys. Rev. {\bf #1}\ (#2)\ #3}
\def\PRL#1#2#3{Phys. Rev. Lett. {\bf #1}\ (#2)\ #3}
\def\ZP#1#2#3{Z. Phys. {\bf #1}\ (#2)\ #3}
\newcommand{\mrm}[1]{\mbox{$\mathrm{#1}$}}
\newcommand{\efmt}[2]{\mbox{$#1 \cdot 10^{#2}$}}
\newcommand{\lum}{\mbox{${\cal L} \,\,\, \rm{cm^{-2}s^{-1}}$}}
\newcommand{\lumunit}{\mbox{$\rm{cm^{-2}s^{-1}}$}}
\newcommand{\eelumi}{\mbox{ ${\cal L}  \, = \, 3 \cdot 10^{33}\,\,\lumunit$}}
\newcommand{\cleolumi}{\mbox{ ${\cal L} \, = \, 2 \cdot 10^{33}\,\,\lumunit$}}
\newcommand{\fnallumi}{\mbox{ ${\cal L} \, = \, 2 \cdot 10^{32}\,\,\lumunit$}}
\newcommand{\btevlumi}{\mbox{ ${\cal L} \, = \, 5 \cdot 10^{31}\,\,\lumunit$}}
\newcommand{\lhcblumi}{\mbox{ ${\cal L} \, = \, 2.0 \cdot 10^{32}\,\,\lumunit$}}
\newcommand{\atlaslumi}{\mbox{ ${\cal L} \, = \, 10^{33}\,\,\lumunit$}}
\newcommand{\lhctoplumi}{\mbox{ ${\cal L} \, = \, 10^{34}\,\,\lumunit$}}
\newcommand{\ra}{\mbox{$\rightarrow$}}
\newcommand{\x}{\mbox{$\cdot$}}
\newcommand{\X}{\mbox{$\times$}}
\newcommand{\degree}{\mbox{$^\circ$}}
\newcommand{\micron}{\mbox{$\mu$m}}
\newcommand{\eg}{\mbox{{\it e.g.}}}
\newcommand{\pcms}{\mbox{$\mathrm{cm^{-2}s^{-1}}$}}
\newcommand{\cms}{\mbox{$\mathrm{cm^2}$}}
\newcommand{\xd}{\mbox{$\mathrm{x_d}$}}
\newcommand{\xs}{\mbox{$\mathrm{x_s}$}}
\newcommand{\Bdl}{\mbox{$\int B$d$l$}}
\newcommand{\chisq}{\mbox{$\chi^{2}$}}
\newcommand{\pt}{\mbox{${\mathrm p_t}$}}
\newcommand{\rpt}{\mbox{$\left| {P\over{T}} \right|$}}
\newcommand{\dbar}{\mbox{$\mathrm  {\overline{d}}$}}
\newcommand{\ubar}{\mbox{$\mathrm  {\overline{u}}$}}
\newcommand{\sbar}{\mbox{$\mathrm  {\overline{s}}$}}
\newcommand{\cbar}{\mbox{$\mathrm  {\overline{c}}$}}
\newcommand{\bbar}{\mbox{$\mathrm  {\overline{b}}$}}
\newcommand{\B}{\mbox{$\mathrm B^{o}$}}
\newcommand{\Bp}{\mbox{$\mathrm B^{+}$}}
\newcommand{\Bm}{\mbox{$\mathrm B^{-}$}}
\newcommand{\Bd}{\mbox{$\mathrm B_{d}$}}
\newcommand{\Bs}{\mbox{$\mathrm B_{s}$}}
\newcommand{\Bu}{\mbox{$\mathrm B_{u}$}}
\newcommand{\Bc}{\mbox{$\mathrm B_{c}$}}
\newcommand{\Bbar}{\mbox{$\rm \overline{B}^{o}$}}
\newcommand{\Bdbar}{\mbox{$\rm \overline{B}_{d}$}}
\newcommand{\Bubar}{\mbox{$\rm \overline{B}_{u}$}}
\newcommand{\Bsbar}{\mbox{$\rm \overline{B}_{s}$}}
\newcommand{\bbbar}{\mbox{$\mathrm {b\overline{b}}$}}
\newcommand{\BBbar}{\mbox{$\mathrm \B\Bbar$}}
\newcommand{\BdBdbar}{\mbox{$\mathrm \Bd -\Bdbar$}}
\newcommand{\BsBsbar}{\mbox{$\mathrm \Bs -\Bsbar$}}
\newcommand{\pipi}{\mbox{$\pi ^+ \pi ^-$}}
\newcommand{\pipipi}{\mbox{$\pi ^+ \pi ^+ \pi ^-$}}
\newcommand{\js}{\mbox{$J/ \psi$}}
\newcommand{\Dz}{\mbox{$\rm D^o$}}
\newcommand{\Dzcp}{\mbox{$\rm D _{CP} ^o$}}
\newcommand{\Dzbar}{\mbox{$\rm \bar{D}^o$}}
\newcommand{\Ds}{\mbox{$\mathrm D_s \!\! ^-$}}
\newcommand{\Dspm}{\mbox{$\mathrm D_s \!\! ^{\pm}$}}
\newcommand{\Dspipipi}{\mbox{\Ds \pipipi}}
\newcommand{\Dsbar}{\mbox{$\rm D_s \!\! ^+$}}
\newcommand{\KKpi}{\mbox{$\rm K ^+ \rm K ^- \pi ^-$}}
\newcommand{\DstoKKpi}{\mbox{$\rm \Ds \, \ra \, \KKpi$}}
\newcommand{\BtomuX}{\mbox{\Bbar \, \ra \' $\mu$ + X}}
\newcommand{\BBbartomu}{\mbox{$\rm B(\Bbar ) \, \ra \, \mu + X$}}
\newcommand{\Btoel}{\mbox{$\rm B \, \ra \, e + X$}}
\newcommand{\BBbartoel}{\mbox{$\rm B(\Bbar ) \, \ra \, e + X$}}
\newcommand{\BdtoJK}{\mbox{$\Bd \, \ra \, \js \, \rm K^o$}}
\newcommand{\BstoJphi}{\mbox{$\Bs \, \ra \, \js \, \phi$}}
\newcommand{\BdtoJKstr}{\mbox{$\Bd \, \ra \, \js \rm K^\star$}}
\newcommand{\BdtoDKstr}{\mbox{$\Bd \, \ra \, \Dzbar \rm K^\star$}}
\newcommand{\Bdtoropi}{\mbox{$\Bd \, \ra \, \rho \pi$}}
\newcommand{\Bdtopipi}{\mbox{\Bd \, \ra \, \pipi}}
\newcommand{\Bdtopipipi}{\mbox{\rm \Bd \, \ra \, \pipi $\pi^o$}}
\newcommand{\Bdtopzpz}{\mbox{\rm \Bd \, \ra \, $\pi^o \pi^o$}}
\newcommand{\BdtoKpi}{\mbox{\rm \Bd \, \ra \, K$^+ \pi^-$}}
\newcommand{\ButoKspi}{\mbox{\rm \Bp \, \ra \, K$^o \pi^+$}}
\newcommand{\BstoKpi}{\mbox{\rm \Bs \, \ra \, K$^- \pi^+$}}
\newcommand{\BstoKK}{\mbox{\rm \Bs \, \ra \, K$^+ K^- $}}
\newcommand{\BstoDspipipi}{\mbox{$\rm \Bs \, \ra \, \Dspipipi$}}
\newcommand{\BstoDsKp}{\mbox{$\rm \Bs \, \ra \, \Ds K^+$}}
\newcommand{\BstoDsKm}{\mbox{$\rm \Bs \, \ra \, \Dsbar K^-$}}
\newcommand{\BstoDspi}{\mbox{$\rm \Bs \, \ra \, \Ds \pi^+$}}
\newcommand{\ButoDK}{\mbox{$\rm \Bp \, \ra \, \Dzbar K^+$}}
\newcommand{\BstomumuKstr}{\mbox{$\rm \Bd \, \ra \, \mu^+\mu^-K^\star$}}
\newcommand{\Bstomumu}{\mbox{$\Bs \, \ra \, \mu^{+} \mu^{-}$}}
\newcommand{\Bdtomumu}{\mbox{$\Bd \ra \mu^{+} \mu^{-}$}}

\newpage

\section{Introduction}

The origin of CP violation in the Standard Model is due to a non-zero complex 
phase in the Cabibbo-Kobayashi-Maskawa~\cite{CKM} (CKM) matrix which describes 
the interaction of left--handed quarks with the charged gauge bosons.
Using the standard phase convention, the two elements of the CMK matrix
acquire large phases; $V_{ub} \sim e ^{-i\gamma}$ and $V_{td} \sim e^{-i\beta}$.
Since these elements involve the 3rd--generation, the
CP violation effects are expected to be large in B systems. The study 
CP--violating effects in B meson decays will allow stringent tests of the 
Standard Model to be made and may point the way to New Physics.

\begin{figure}[ht]
\begin{center}
\end{center}
\mbox{\epsfig{file=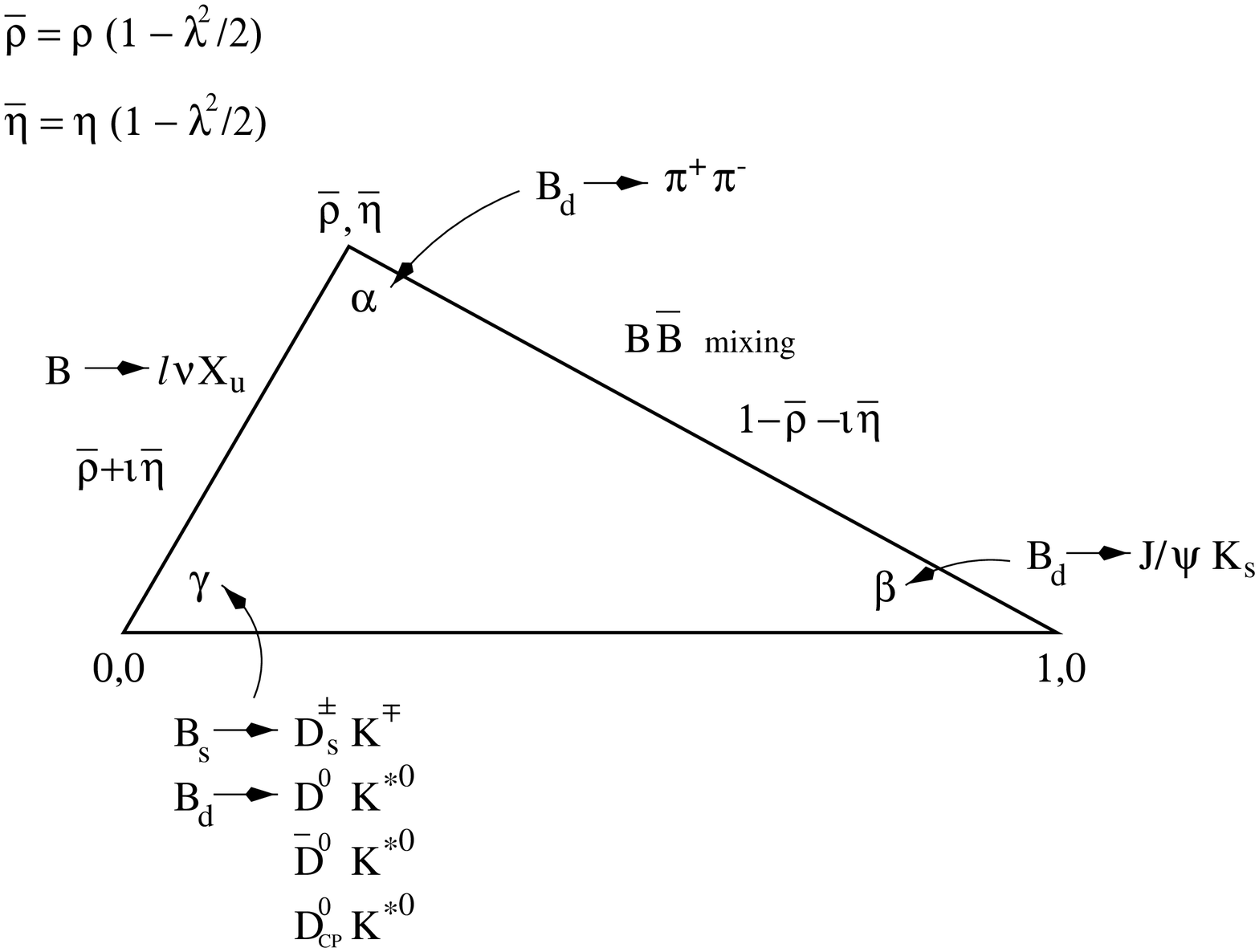,width=\columnwidth}}
\caption[]{Unitarity Triangle in $\overline{\rho.}, \overline{\eta}$ plane 
defining the angles $\alpha, \beta$ and $\gamma$. The 
preferred decay channels used to extract the parameters of the 
Unitarity Triangle is also shown. }
\label{fig:unitarity}
\end{figure}

Unitarity of the CKM matrix implies  triangular relations such as
\begin{equation}
V_{ud}V^*_{ub} + V_{cd}V^*_{cb} + V_{td}V^*_{tb} = 0.
\end{equation}
Figure~\ref{fig:unitarity} shows the so--called Unitarity Triangle using the 
Wolfenstein parametrization~\cite{wolfenstein}. In drawing the triangle, 
$V^*_{ud} \approx 1$, $V_{tb} \approx 1$ and $V_{cs} \approx -\lambda$ are 
used, and the sides of the triangle are scaled by $A\lambda ^3$. We already 
have some information on the magnitudes of the sides of the Unitarity Triangle 
from present experiments. For example;
\begin{itemize}
\item $b \ra c$ transitions in inclusive and exclusive B meson decays
provide information on $|V_{cb}| = A\lambda^2$ and set the scale of the 
Unitarity Triangle, 
\item $|V_{ub}/V_{cb}|$, which is related to 
the length of the left side of the 
Unitarity Triangle, is obtained from $b \ra u$ transitions in charmless B 
meson decays, 
\item $|V_{td}|$ is extracted from $\B \leftrightarrow \Bbar$ 
oscillations. 
\end{itemize}

The primary goal of future B experiments (including the ones being upgraded) 
is to measure both the sides and the angles of the Unitarity Triangle and by 
over--constraining it, test the consistency of the Standard Model.

The study of Flavor--Changing--Neutral--Currents (FCNC) in B decays also 
provides quantitative information on the CKM matrix--elements, 
$V_{td}, V_{ts}$ and $V_{tb}$.  In the Standard Model, FCNC are allowed only 
through loop (box and penguin) diagrams and are therefore strongly suppressed. 
If contributions from New Physics beyond the Standard Model are comparable in 
magnitude to loop diagrams, FCNC decays could provide one of the earliest 
routes to discover such physics. 
  
An extensive experimental program is underway to study CP violation in B meson 
decays.  There is at least one experiment under construction or planned at 
every high energy physics laboratory, as summarized in Table ~\ref{bprogram}. 
The Beauty97 conference provided a forum for these experiments to discuss their
physics potential and experimental challenges relating to these studies. The 
status and future plans of these projects, as well as the recent results from 
key R\&D projects on various sub--detectors were presented.
In addition, there were presentations on new B physics results from ongoing 
experiments and theoretical talks on the latest developments in the field.
The recent advances in theoretical B physics discussed at the conference
were summarized by Jon Rosner~\cite{rosner}. Therefore, in this paper, I only
give the summary of the experimental B physics program. 
Readers should refer to individual contributions at these proceedings for 
further details.

\section{General remarks}

The Standard Model description of CP--violation and the formalism used to 
extract some of the CKM matrix elements from B meson decays is found throughout
the literature (see, for example, a recent review by 
R. Fleischer~\cite{fleischer} and the references therein). Experiments usually
make simplifying assumptions when applying these formalisms to demostrate their
physics potential. Without explicitly examining these simplifications, some of
which may not always apply, we can not understrand the measurements. Here we 
comment on some of the analysis simplifications and limitations in the B 
program.

\subsection{$\boldmath \B \leftrightarrow \Bbar$ oscillations}

Second--order weak--interactions involving box diagrams allow $\B  
\leftrightarrow \Bbar$ transitions. If one ignores CP violation, the
resulting decay distributions for a \B\ produced at $t = 0$ , and decaying
at time $t$ into flavor--specific final states, \B\ or \Bbar ,
corresponding to non--oscillation and oscillation, respectively, are:
\begin{equation}
R(\B _{t=0} \, \ra \, \B(t)) \, \sim \, 
e^{-\Gamma t} \, [cosh({\Delta \Gamma \over{2}} t) + cos(\Delta mt)]
\label{noosci}
\end{equation}
\begin{equation}
R(\B _{t=0} \, \ra \, \Bbar(t)) \, \sim \, 
e^{-\Gamma t} \, [cosh({\Delta \Gamma\over{2}} t) - cos(\Delta mt)]
\label{osci}
\end{equation}
\noindent
where t is the proper time, $\Gamma$ is the average decay width of the two mass
eigenstates and $\Delta \Gamma$ and $\Delta m $ are the decay width difference 
and the mass difference between the heavy and light B mesons, respectively.
Similar decay distributions exist for \Bbar . 

In order to observe the oscillations in the two decay distributions, the flavor
of the B meson at production has to be established (referred to as flavor 
tagging). Since, for various reasons, flavor tagging is imperfect, the 
experimentally observed decay time distributions are mixtures of the two ideal 
distributions, Eq. \ref{noosci} and \ref{osci}:
\begin{equation}
R_{obs}(\B _{t=0} \, \ra \, \B(t)) \, \sim \, A(t) \, 
e^{-\Gamma t} \, [1 + D cos(\Delta mt)]
\label{eq:noosci}
\end{equation}
\begin{equation}
R_{obs}(\B _{t=0} \, \ra \, \Bbar(t)) \, \sim \, A(t) \, 
e^{-\Gamma t} \, [1 - D cos(\Delta mt)]
\label{eq:osci}
\end{equation}

\noindent
where, for simplicity, we assume $\Delta \Gamma = 0$ (a good approximation for
\Bd , but not for \Bs). $A(t)$ is the experimental acceptance and D is the 
dilution factor, defined as $\rm D = {{N_g - N_b}\over{N_g + N_b}}$ with 
$\rm N_g$ and $\rm N_b$ are the numbers of good and bad tags, respectively.  

\begin{figure}[ht]
\begin{center}
\mbox{\epsfig{file=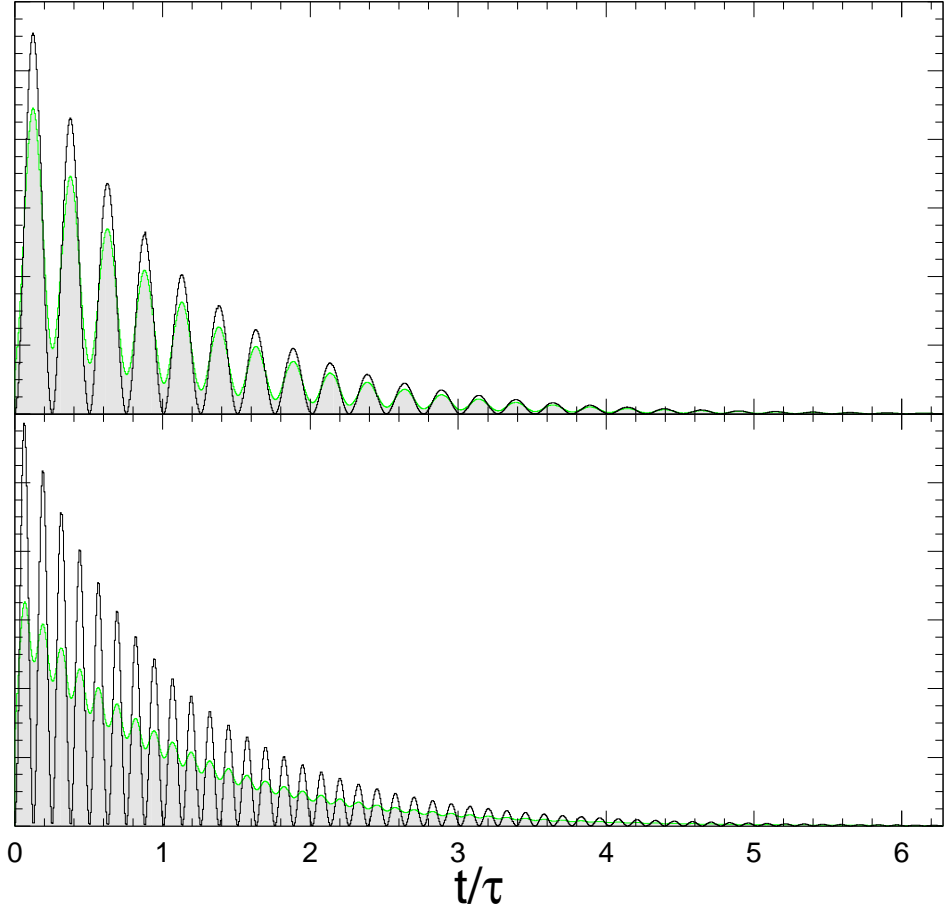,width=\columnwidth}}
\end{center}
\caption[]{Expected decay time distribution for oscillated \Bs\ decays 
(\Bs \ra\ \Bsbar \ra X) for \xs = 25 and 50. Solid curves are the ideal
(no dilution and perfect vertex (time) resolution) decay time distributions.
Shaded curves show the effect of a vertex resolution ($\sigma _t = 0.04 ps$).}
\label{fig:tres}
\end{figure}

For \Bs\ oscillations, the vertex resolution, and hence the proper time 
resolution, limits the experiments ability to observe very large $\Delta m_s$ 
values.  Figure~\ref{fig:tres} shows the effect of a finite vertex resolution 
for two values of $\xs =\Delta m_s /\Gamma = $ 25 and 50, assuming ideal 
tagging, D=1.  The solid curves are the ideal decay distributions, 
Eq.~\ref{eq:osci}, for B decays after oscillation, with $A(t) = 1$.
The shaded functions are the same distributions smeared with a proper time
resolution of 4\%. The effect of the finite proper time resolution is to damp 
the amplitude of the oscillations similar to a large dilution (small D). 
In this case, the measured amplitude of the oscillations would be the 
combined effects of the dilution and the proper time resolution. 

Although the frequency of very small amplitude oscillations could 
be measured using a large sample of events (collected over many years), 
in practice, the observation of the oscillations will be limited to \xs\ 
values of $\xs < {3-4  \over{\sigma_{\tau}/\tau}}$.

\subsection{Angles of the Unitarity Triangle}
The most straightforward 
way to measure the angles $\alpha$ and $\beta$ of the Unitarity 
Triangle is to measure the asymmetry between \Bd\ and \Bdbar\ decays to a CP 
eigenstate, $f$. In this case, the time--dependent asymmetry is given by:
\begin{eqnarray}
A(t) & = &
{{R(\Bd (t) \ra f) - R(\Bdbar (t) \ra f)}\over{R(\Bd (t) \ra f) 
+ R(\Bdbar (t) \ra f)}} \nonumber \\
\nonumber \\
& = & A^{dir}_{CP} \, cos(\Delta mt) + A^{mix}_{CP} \, sin(\Delta mt)
\label{assymmetry}
\end{eqnarray}
where the amplitude $A^{dir}_{CP}$ and $A^{mix}_{CP}$ of the $cos$ and $sin$ 
terms are the $direct$ and the $mixing$--$induced$ CP--violating contributions. 
In Eq.~\ref{assymmetry}, \B\ and \Bbar\ refer to the flavor of the B meson at 
production which must be identified by means of tagging.

If only a single phase dominates the decay, as is the case for \BdtoJK , then
\[A^{dir}_{CP} \, = \, 0\] 
\[A^{mix}_{CP} \, = \, \xi \, sin(\Phi_M - \Phi _D) \]
where $\xi$ is the CP eigenvalue of $f$, 
and $\Phi_M$ and $\Phi_D$ are the mixing and decay phases, respectively.
If there is no contribution other than the Standard Model to
$\Bd \leftrightarrow \Bdbar$ mixing, then $\Phi_M = 2\beta$. 
Experiments usually assume $A^{dir}_{CP} = 0 $ and quote the error on 
$A^{mix}_{CP}$ as a measure of their CP reach. 

Since tagging is imperfect, the observed $A^{obs}_{mix}$ will be the 
product of the dilution D and $A^{mix}_{CP}$.
Therefore, contrary to the oscillation studies, one needs to know the 
magnitude of the dilution from independent measurements
in order to extract the true value of $A^{mix}_{CP}$.
Experiments always assume that the backgrounds have no 
CP asymmetry and are treated as an additional dilution to $A^{mix}_{CP}$.
Although this assumption would be valid for combinatoric backgrounds, it is 
not correct for 2--body background under the \Bdtopipi\ which might have its
own CP asymetry. 

\paragraph{The Angle $\boldmath \beta :$}
Everybody's favourite decay mode for measuring the angle $\beta$ is \BdtoJK . 
In this decay mode, the dominant decay diagram is the $\bbar \ra \cbar c \sbar$
tree transition with $\Phi_D = 0$ 
(the penguin contributions are expected to be small and have the same 
phase as the tree amplitude), yielding to good approximation:
\[A^{dir}_{CP} \, = \, 0 \,\,\,\, \, \, {\rm and} 
\, \, \,\,\,\,A^{mix}_{CP} \, = \, -sin(2\beta) \]

\paragraph{The Angle $\boldmath \alpha :$}
The preferred method of measuring the angle $\alpha$ is through the decay mode, 
\Bdtopipi . 
In fact, this decay mode measures the sum of the angles $\gamma$ and $\beta$ .

Again if one totally ignores the penguin pollution, since the phase 
contributed by the dominant $\bbar \ra \ubar u \dbar$ decay is 
$\Phi_D = -2\gamma$, 
\[A^{mix}_{CP} = sin(2(\gamma + \beta) = -sin(2\alpha)\]
\noindent
Here we used the triangular relation $\alpha = \pi - \gamma - \beta$.

The presence of QCD penguins with phases different than the leading tree 
contribution in this decay mode leads to $direct$ CP violations, 
$A^{dir}_{CP} \neq 0$.
In this case
\[A^{dir}_{CP} = 2 \, \rpt \, sin(\Delta ) \, sin(\alpha) \]
\[A^{mix}_{CP} = -sin(2\alpha) 
- 2 \, \rpt \, cos(\Delta) \, cos(2\alpha) \, sin(\alpha) \]

\noindent
where \rpt\ is the ratio of penguin to tree contributions and $\Delta$
is the strong phase difference between the penguin and tree diagrams.
Even if $direct$ CP violation turns out to be too small to observe, 
the correction to the simple assumption, 
$A^{mix}_{CP} = -sin(2\alpha)$, could be very large.

In principle, the effect of penguin pollution could be 
eliminated by measuring the branching ratios of
$BR(\rm B^+ \ra \pi^+\pi^o), \, \, BR(\Bdtopzpz )$ and their charge
conjugate channels~\cite{gronaulondon}. 
But, due to the small $BR(\Bdtopzpz )$, 
this method would be difficult to realize from the experimental point of view.

If \rpt\ is known, $\alpha$ and 
$\Delta$ can be extracted from the observed asymmetry.

\paragraph{The Angle $\boldmath \gamma $ Method--1:}

One way to measure the angle $\gamma$
is to use \Bs\ decays into $\rm D_s^{\pm} \rm K^\mp$~\cite{aleksan}.
Here, the decay time distributions depend on both 
the weak phase, $\gamma $ and the strong phase, $\delta$. 
$sin(\gamma + \delta)$ and $sin(\gamma - \delta)$ 
can be obtained by fitting to four decay time distributions of \Bs\ 
and \Bsbar , with and without oscillations. Here we assume 
that $V_{ts}$ is real. In fact, the weak phase measured with this method is
$\gamma - 2 \delta '$, where $\delta '$ is the phase of $V_{ts}$. 
The estimated error 
on the angle $\gamma$ depends on the values of $\xs , \gamma$ and $\delta$. 

\paragraph{The Angle $\boldmath \gamma $ Method--2:}

\begin{figure}[ht]
\begin{center}
\mbox{\epsfig{file=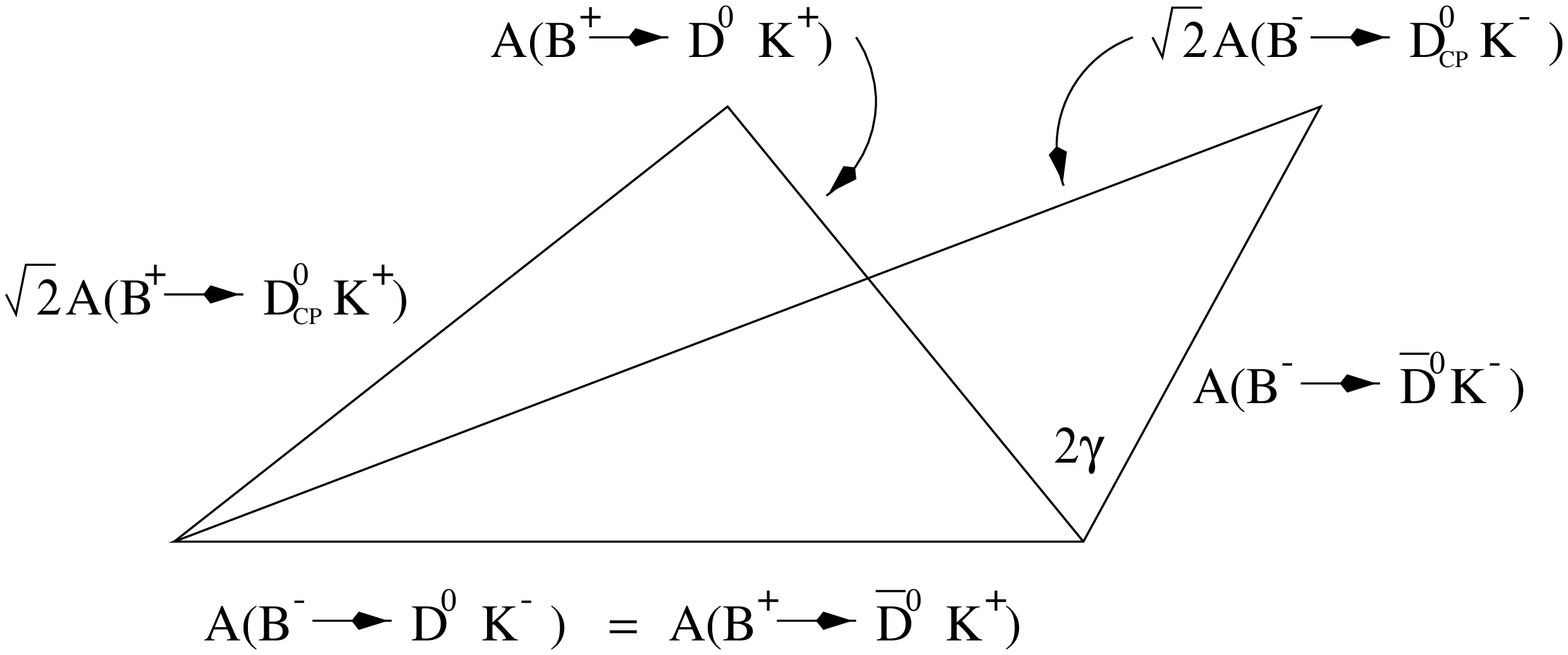,width=\columnwidth}}
\end{center}
\caption[]{Triangular relation between the decay amplitudes of 
B$^+$ \ra\ \Dzcp K$^+$ ,
B$^+$ \ra\ \Dzbar\ K$^+$ and 
B$^+$ \ra\ \Dz\ K$^+$
and their charge-conjugate states, where 
$\Dzcp=(D^o+\overline{D}^o)/\sqrt{2}$.}
\label{fig:method-II}
\end{figure}

A second method of measuring $\gamma$~\cite{gronauwyler} 
consists of measuring the exclusive decay rates in the following 3 channels:
B$^+$ \ra\ \Dzcp K$^+$ ,
B$^+$ \ra\ \Dzbar\ K$^+$ and 
B$^+$ \ra\ \Dz\ K$^+$
and their charge--conjugate states, where 
$\Dzcp=(D^o+\overline{D}^o)/\sqrt{2}$.
The angle, $\gamma$, is then extracted by forming two triangles using these six 
amplitudes in the complex plane, as shown in Fig.~\ref{fig:method-II}.
Unfortunately, these triangles are very flat; therefore, this method is 
experimentally very challenging.

Analogous channels exist for \B\ decays~\cite{dunietz} where the corresponding
triangles are not as flat as for charged B's. 

\subsection{Tagging strategies}

\begin{figure}[ht]
\begin{center}
\mbox{\epsfig{file=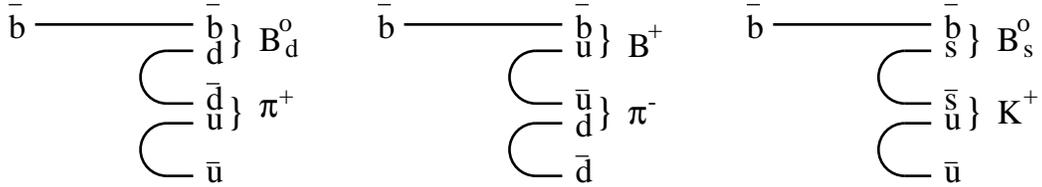,width=\columnwidth}}
\end{center}
\caption[]{Flavor diagram showing the correlation between the flavor and charge
of the B meson and the charge of near by $\pi$ or K produced in the 
fragmentation.}
\label{fig:Ftag}
\end{figure}

As explained above, many B physics studies require knowledge of the flavor of 
the neutral B--meson at its time of creation. This information is usually 
obtained by examining the decay products of the accompanying \Bbar\ meson. 
Since flavor tagging is not always correct, the purity of the tag, as well as
its efficiency, plays an important role in extracting the interesting 
physics quantities from the decay time distributions.  A figure of tagging 
merit is the product of the tagging efficiency, $\epsilon$, and the square of 
the dilution, D. The most common tagging methods are:
\begin{itemize}

\item High--\pt\ leptons from semileptonic B meson decays. 
This type of tagging uses the correlation between the charge of the 
lepton and the flavor of the b quark. Selection of 
high--\pt\ leptons decreases the contamination from charm and $K/\pi$ decays. 
Although in this mode the tag is very pure, due to the small branching ratio of 
semileptonic B decays, this tag is relatively inefficient.
Nevertheless, high--\pt\ leptons constitute the main tagging
particles for hadronic B final states which are triggered by the tagging 
lepton.

\item Charged kaons from $b \, \ra \, c \, \ra \, s$ transitions. 
This method relies on the correlation of the kaon charge 
with the b quark flavor. In order to enhance the puritiy of kaons from
b decays, a large impact parameter to the primary vertex or association of the 
kaon to a secondary vertex is required. This tag is relatively 
efficient, but it suffers from the badly measured (multiple scattered) 
kaons along with the wrong--sign kaons from $\rm D_s$ decays. 
The rapid oscillations of \Bs 's make the kaons and leptons from these 
decays useless.

\item Jet--charge tags. This methods relies on the correlation between the
 b quark flavor and its charge which is inferred from the total charge of the 
b--jet. Since it is relatively hard to associate which of the primary tracks 
belongs to a b--jet, the jet charge is calculated as a weighted sum of all 
track charges inside a cone around the B meson. This type of tagging has been 
successfully employed at LEP~\cite{LEP} and could be applied both to the jet 
containing the constructed B meson (same--side) and the accompanying 
\bbar -jet (away-side).

\item Fragmentation tags. Figure~\ref{fig:Ftag} demonstrates the correlation 
between the nearby particle's charge, produced in the fragmentation process, 
and the \bbar\ quark flavor.  During the fragmentation of \bbar\ quarks to 
\Bd , a d quark is picked up from the vacuum, leaving a \dbar\ in the vicinity 
of the \Bd\ meson. If this \dbar\ forms a charged pion, in this case  a \Bd\ is 
always associated with a $\pi^+$. Similarly, \Bu\ is always associated with a 
$\pi^-$. Therefore, in principle, the flavor of the B meson could be tagged
by studying the charge of the pion close to the reconstructed B meson. 
Since the charge--flavor correlation for the \Bd\ and \Bu\ mesons are opposite,
in order for this tag to function with high purity, the charge of the B meson 
should be correctly identified. This method is less attractive for semileptonic
decays, since it is very easy to miss a slow pion and misidentify the B meson 
charge, hencedirty introducing large dilutions. This type of tagging is 
employed by the CDF collaboration in their oscillation analysis~\cite{Bauer}.      

Similar correlations exist between the \Bs\ flavor and the charge of a 
nearby kaon from the primary vertex~\cite{p238}.
\end{itemize}
 
\section{New results}

While we await the discovery of CP violation in B mesons decays,
a wealth of new information on other B physics topics
is coming from CLEO, CDF, D\O\ and the LEP experiments.
Here I summarize some of these results presented at this meeting
which have direct
effects on our understanding and expectations from the future B experiments.
\begin{itemize}
\item Many measurements of the \Bd\ oscillation frequency from the LEP, CLEO 
and ARGUS experiments yield the world average~\cite{LEP}: 
\[\Delta m_d \, = \, 0.463 \pm 0.018 \, \,ps^{-1} \]

\item CDF presented\cite{Bauer}: 
$\Delta m_d = 0.474 \pm 0.029 \pm 0.026 \,\,ps^{-1}$, which is consistent with 
the weighted average from the $e^+e^-$ experiments. The CDF error is comparable
to the measurement errors of the individual $e^+e^-$ measurements, thus 
demonstrating that precise B physics measurements can be carried out in a 
``dirty" hadron environment.

\item The combined results on \Bs\ oscillations from LEP\cite{LEP} 
yield a lower limit for $\Delta m_s$. 
\[\Delta m_s > 10.2 \,\, ps^{-1} \rm \,\, at \,\,95\% \,\,C.L.\]
LEP will not be taking new data at the Z$^0$ mass, but the 
$\Delta m_s$ sensitivity could still be improved with new analysis techniques
of the existing data. SLD is also expected to have an impact with their precise
vertex detector. The combined $\Delta m_s$ sensitivity could probably lie in 
the range 15--20 $ ps^{-1}$ \cite{LEP}. The existing lower limit already 
narrows the discovery window of the \Bs\ oscillations at the 1st--generation 
hadron B experiments\footnote{No \Bs 's are produced at $e^+e^-$ machines 
operating at the $\Upsilon _{4S}$.}. On the other hand, if 
$\Delta m_s  > 15 \,\, ps^{-1}$, it would be possible to observe the lifetime 
differences between the heavy and light \Bs 's directly.

\item Both CDF and D\O\ reported that the \bbbar\ cross sections are 
significantly larger than one expects from NLO QCD calculations, although the 
shape of the \pt\ spectrum is in reasonable agreement with the theory. However,
the s--dependence of the cross section agrees with the QCD 
predictions~\cite{Bauer,Baarmand}. 
Although this is good news for the future LHC B--experiments (larger 
production rate), it also points out a deficiency in our understanding of 
heavy quark production.

\item In the Flavor--Changing--Neutral--Current sector, CDF reported new 
upper limits for $BR(\Bdtomumu ) < 6.9 \X 10^{-7}$ and 
$BR(\Bstomumu ) < 2.1 \X 10^{-6}$~\cite{Bauer}. 
These limits are two orders--of--magnitude larger than Standard Model 
expectations. 

\item Many new branching ratios and/or upper limits for several classes 
of rare decays are reported by the CLEO collaboration~\cite{Alan}. 
From these, $BR(\BdtoKpi) =
(1.5^{+0.5}_{-0.4} \pm 0.1) \X 10^{-5}$ and $BR(\ButoKspi) = (
2.3^{+1.1}_{-1.0} \pm 0.2) \X 10^{-5}$ already generated considerable 
excitement since, in principle, such measurements (with smaller errors)
will place constraints on possible values of the angle, $\gamma$.

\item Even though CLEO only reports an upper limit for $BR(\Bdtopipi) < 
1.5 \X 10^{-5}$, using the ratio of $N_{\pi \pi} / N_{K\pi}$ from their 
combined maximum--likelihood fits to B \ra $hh$, one could calculate 
$BR(\Bdtopipi) = 7 \X 10^{-6}$. If confirmed, this would make the 
observation of CP violation in the \Bdtopipi\ decay mode very unlikely at the
1st--generation  experiments. 
\end{itemize}

\section{First generation CP--violation B--experiments}

In the five experiments preparing to study CP violation in B decay, there
are three distinctly different pioneering approaches.
\begin{itemize}

\item Experiments in asymmetric $e^+e^-$ machines: BABAR~\cite{hitlin,BABAR}
 and BELLE~\cite{bondar,BELLE},

\item General purpose high--\pt\ (central) experiments: CDF\cite{nigel} 
and D\O\ \cite{wayne},

\item A fixed--target experiment at a proton storage ring: 
HERA-B~\cite{lohse,HERAB}.

\end{itemize}
While all these experiments have similar CP--reach in the ``gold--plated'' 
channel, \BdtoJK , the three approaches forms a complementary set
with their different systematic uncertainties and  various strengths.

\subsection{Asymmetric $e^+e^-$ experiments: BABAR \& BELLE}

\begin{figure}[ht]
\begin{center}
\mbox{\epsfig{file=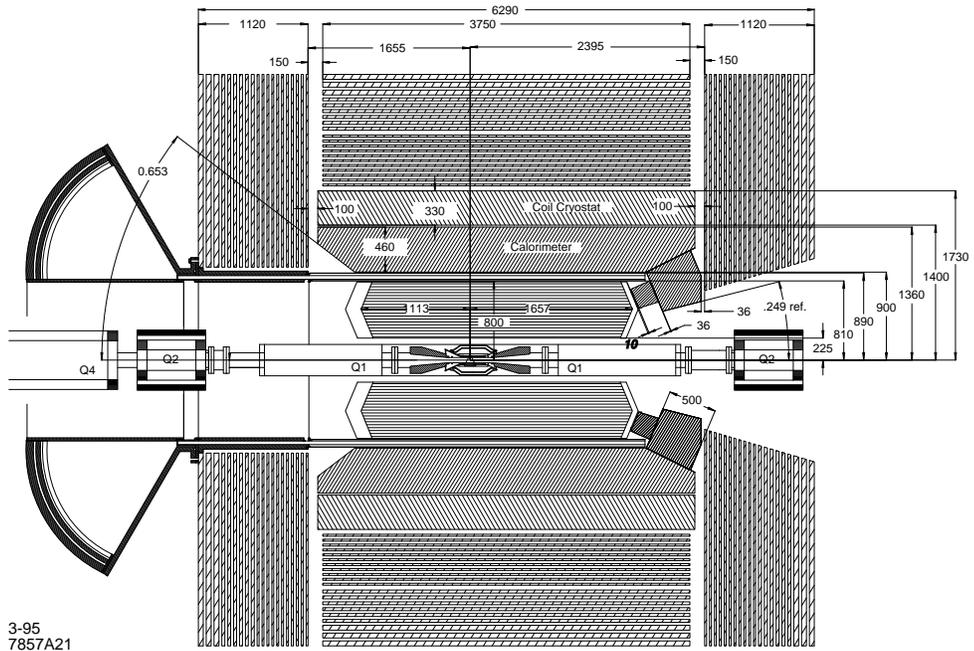,width=16cm}}
\end{center}
\caption[]{Overall view of the BABAR spectrometer.}
\label{fig:BABAR}
\end{figure}

\begin{figure}[ht]
\begin{center}
\rotatebox{-90}{\mbox{\epsfig{file=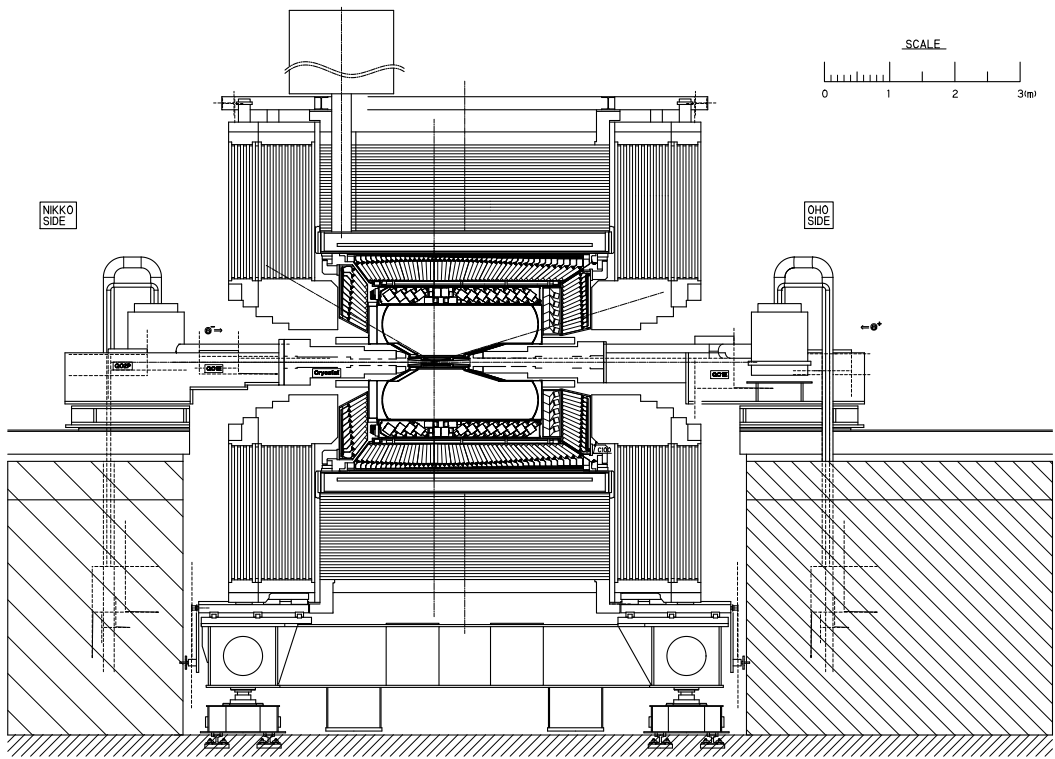,width=10cm}}}
\end{center}
\caption[]{Overall view of the BELLE spectrometer.}
\label{fig:BELLE}
\end{figure}

Figures~\ref{fig:BABAR} and \ref{fig:BELLE} shows the detector layouts of the 
$e^+e^-$ experiments under construction, BABAR at SLAC and BELLE at KEK, 
respectively. Both experiments are very similar in their detector designs, 
differing mainly in their approach to particle identification. While the BELLE 
experiment uses a combination of time--of--flight detectors and Aerogel 
Cherenkov counters, the BABAR experiment relies on a new 
type of Ring--Imaging--Cherenkov counter, DIRC~\cite{dirc}.
The DIRC detector uses quartz bars as a radiation medium; the produced 
Cherenkov light is transported using internal reflections in the radiators,
to photon detectors located outside the magnetic volume.
The construction of both detectors are proceeding according to schedule;
they are both expected to become operational in the second half of 
1999~\cite{hitlin,bondar}.
 
\subsection{General purpose high--\pt\ experiments: CDF \& D\O }

The existing central high--\pt\ experiments, CDF and D\O, at the Tevatron 
collider are undergoing major upgrades to prepare for RUN--II. Although the 
main purpose of the upgrades is to prepare the detectors for the luminosity 
increase with the new Main Injector, many of the improvements will also 
increase the B--physics reach of the experiments. Of course, the B program also 
benefits from the increased luminosity of RUN II.  

\begin{figure}[ht]
\begin{center}
\mbox{\epsfig{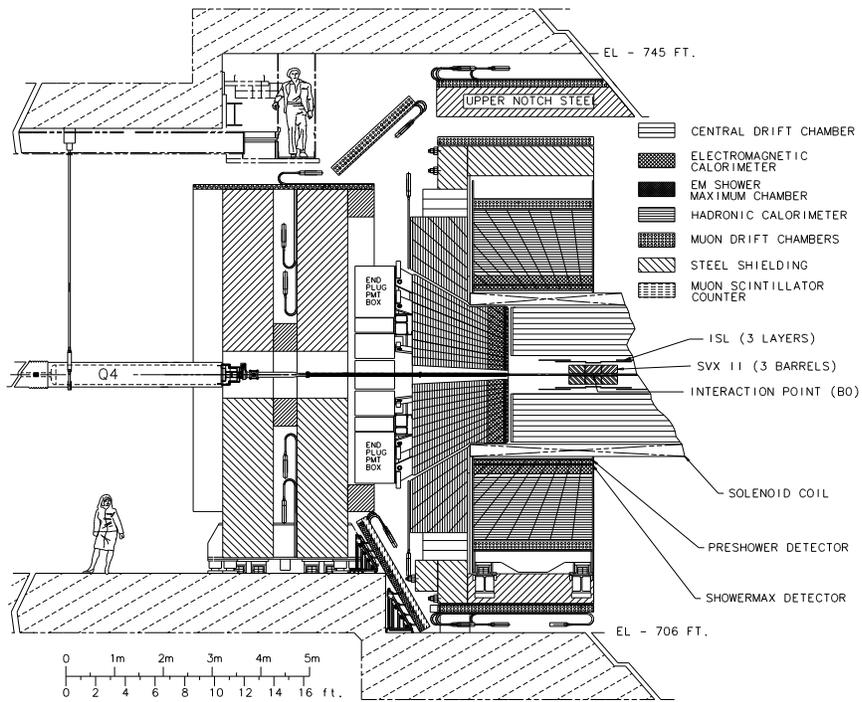}}
\end{center}
\caption[]{Elevation view of half of CDF II detector.}
\label{fig:CDF}
\end{figure}

Figure~\ref{fig:CDF} shows half of the CDF II  detector layout (after upgrade).
A new Silicon--Micro--Vertex detector of CDF II will increase its vertex 
acceptance and resolution. In addition, a new second level trigger will
allow CDF II to trigger on hadronic B decays, opening up the possibility of 
using decay modes other than ones containing \js , as has been done until now.

\begin{figure}[ht]
\begin{center}
\mbox{\epsfig{file=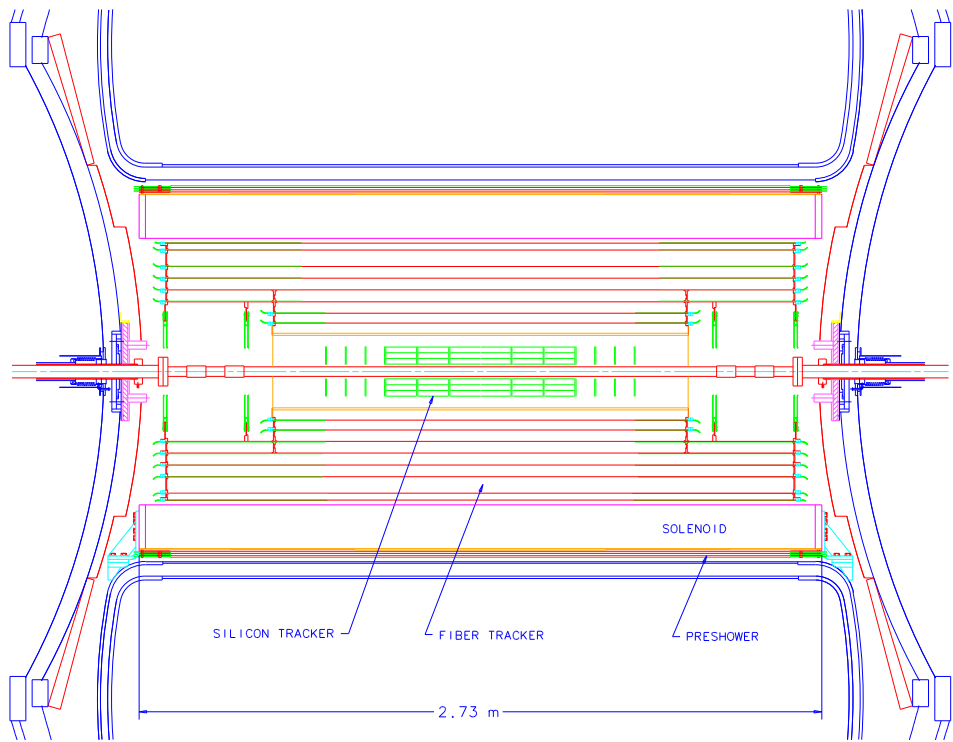,width=\columnwidth}}
\end{center}
\caption[]{The schematic view of the D\O\ tracker system and the new 
solenoid magnet.}
\label{fig:D0}
\end{figure}

The D\O\ upgrade is more like a rebuild. Although they will retain their 
excellent calorimetry, they are adding a solenoid magnetic field and  
a silicon microvertex detector along with an entirely new 
tracking system. Figure~\ref{fig:D0} shows the layout of 
their new tracking system and the magnet.

Both experiments will start data taking again in 2000~\cite{nigel,wayne}. 

\subsection{Fixed--target experiment: HERA-B}

\begin{figure}[ht]
\begin{center}
\mbox{\epsfig{file=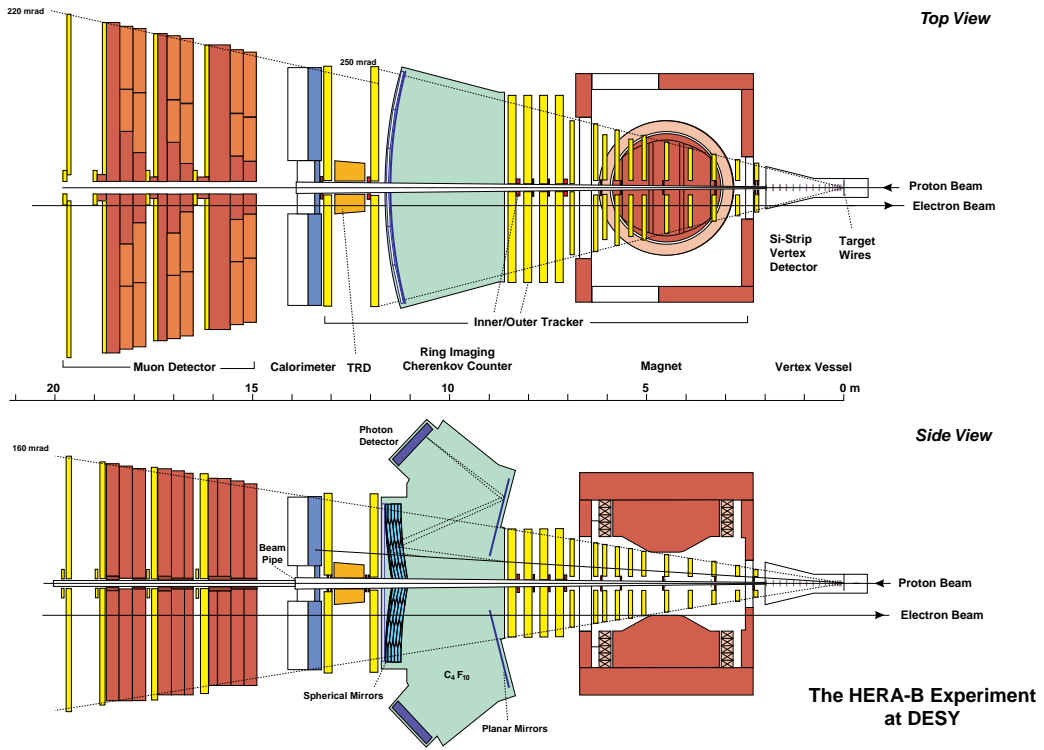,width=\columnwidth}}
\end{center}
\caption[]{Plan and elevation views of the HERA-B spectrometer.  }
\label{fig:HERAB}
\end{figure}

Hera-B is a fixed--target experiment which uses the stored 820 GeV HERA 
proton beam~\footnote{There are plans to increase the beam energy 
above 900 GeV.} and a non--invasive internal fixed--target consisting of 8 wires.
Figure~\ref{fig:HERAB} shows the layout of the HERA-B spectrometer with the 
proton beam pipe going through the detector. Wire targets are placed in the 
same conical vacuum tank which houses the silicon vertex 
detector~\cite{kirsten}. The RICH detector behind the magnet provides kaon 
identification over the full momentum range~\cite{rosen}. The main Level--1 
trigger of HERA-B is a pair of high--\pt\ leptons or hadrons with cuts on the 
invariant mass of the pair (\js\ mass in case of dilepton trigger and B mass 
in case of high--\pt\ dihadron trigger)~\cite{mad}. 
 
The rather small 
\bbbar\ cross section not far above \BBbar\ threshold implies that
the HERA-B detector has to operate with interactions rates of 40 MHz, 
corresponding to 4 superimposed interactions per proton bunch crossing.
The HERA-B detector has to cope with the correspondingly high particle 
densities and the data acquisition system has to handle high data flows, 
similar to those in the LHC experiments.

The experiment had successful engineering runs in 1996 and 1997, with prototype
and/or final detector modules of each subsystem. During the 1997 run, the wire 
targets routinely operated with 40 MHz interaction rates and all detectors were
read out simultaneously through the common data acquisition system.

Operational problems in the harsh high--rate hadron environment encountered 
in the (inner tracker) Micro--Strip--Gas Chambers (see discussion of 
solution in Sec.~\ref{sec:msgc}) and the (outer tracker) honeycomb drift 
chambers has delayed the startup of full HERA-B operation. Major efforts are 
underway to find a solution to the use of honeycomb drift chambers in the 
HERA-B environment and to complete as much as possible of the tracking system 
for the 1999 running period~\cite{lohse}. Independently, a back--up scenario 
for the outer tracker\cite{lohse}, based on (ATLAS TRT) straw tubes of the type
used in the HERA--$B$ transition radiation detector\cite{saveliev} which has 
survived HERA--$B$ high--rate runs without problems, is being worked out.

\subsection{Symmetric $e^+e^-$ experiment: CLEO upgrade}

The CLEO experiment is preparing for the next luminosity upgrade of the CESR 
$e^+e^-$ storage ring, where the luminosity of \cleolumi\  will be reached.  
In addition to changes to the inner tracking volume to house the new machine 
elements, four new layers will be added to the silicon vertex detector.
The most important addition will be the RICH detector module with LiF 
radiator. The new RICH detector will provide the critical $\rm K/\pi$ 
separation needed for rare 2--body B decays~\cite{CLEOIII}.

\section{Future B--experiments}

\subsection{General purpose high--\pt\ experiments: ATLAS \& CMS}

The two general purpose high--\pt\ experiments, ATLAS~\cite{atlas} and 
CMS~\cite{cms}, planned for the LHC at CERN also have extensive B physics 
programs which exploit the large \bbbar\ cross section at these energies.

The excellent muon systems~\cite{ATLASmuon,CMSmuon} and electromagnetic 
calorimeters~\cite{emcal} of ATLAS and CMS will allow them to trigger on and 
study B mesons containing \js\ or leptons in the final state. 
The pixel detextors in the most inner part of the 
tracking systems of ATLAS~\cite{damon,stephen} and CMS~\cite{muller,manola}, 
have a very good vertex resolution which are critical for B physics.
Due to the difficulty of reconstruction and tagging of B mesons 
in the presence of multiple interactions in the same bunch crossing, 
most of the B physics program in these experiments will be limited to 
the early days of the LHC when the machine luminosity will be about 
$10^{33}$~cm$^{-2}$~s$^{-1}$. 

The main trigger for these experiments is a high--\pt\ lepton(s). 
Even if one learns to do B physics with multiple interactions, 
the limited bandwidth of the Level--1 trigger output will force them to run 
with even higher than the already high \pt\ thresholds (except for the
dimuon trigger), thus limiting 
the usefulness of the higher luminosities available at LHC for B physics.

Most of the Technical Design Reports from both experiments are already
approved and 
construction of the various subsystems is starting~\cite{tartarelli,muller}.

\subsection{The ultimate B experiments: BTeV \& LHCb}

It is generally accepted that an experiment optimized to exploit the 
large \bbbar\ cross sections in the forward directions at 
hadron colliders will be 
the ultimate B experiment in any foreseeable future. 
Both the BTeV proposal~\cite{penny,BTEV} for the Tevatron and 
the LHCb proposal~\cite{neville,LHCB} for LHC are in this class.

Due to the forward peaking of \bbbar\ production, both experiments have a 
forward geometry\cite{schlein} similar to fixed--target experiments. 
Both experiments
could use similar designs and technologies in many of their subsystems. 
The fundamental differences are limited to their magnetic configurations, 
silicon vertex detectors and trigger strategies. 

\begin{figure}[ht]
\begin{center}
\rotatebox{-90}{\mbox{\epsfig{file=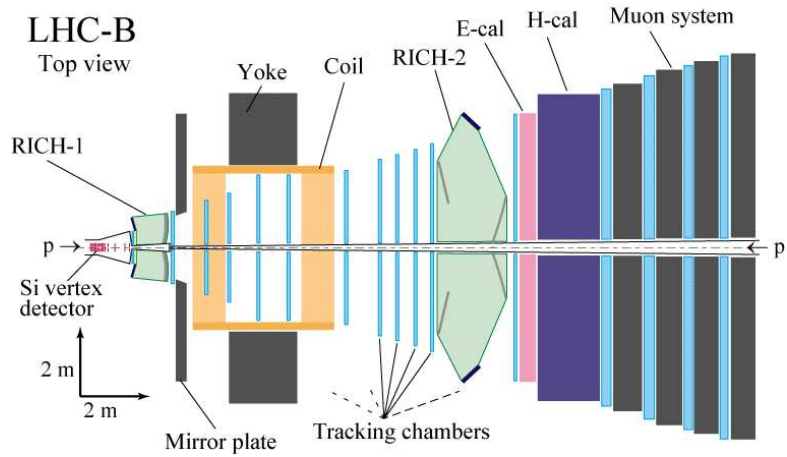,width=6cm}}}
\end{center}
\caption[]{Layout of the LHCb spectrometer. 
}
\label{fig:LHCB}
\end{figure}

\begin{figure}[ht]
\begin{center}
\mbox{\epsfig{file=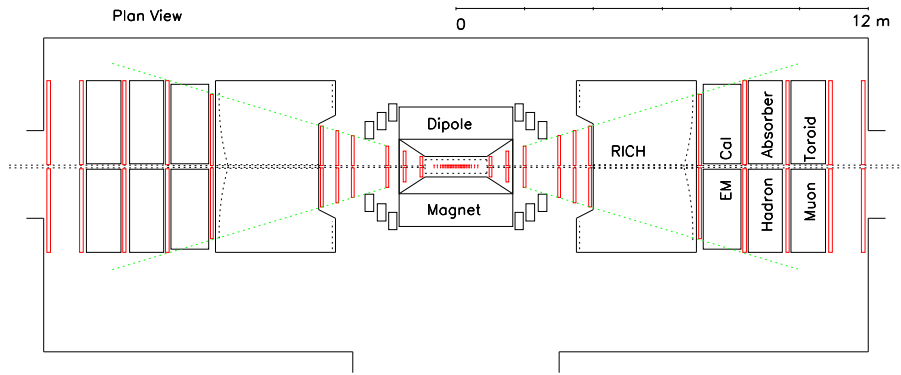,width=\columnwidth}}
\end{center}
\caption[]{Layout of the BTeV two-arm spectrometer.
}
\label{fig:BTEV}
\end{figure}

Figure~\ref{fig:LHCB} shows the layout of the proposed LHCb spectrometer.
The LHCb detector is a single--arm spectrometer with a 4.2 Tm dipole magnet 
centered at 4m from the interaction region. 
In contrast, BTeV is a 2--arm spectrometer with a single dipole magnet over 
the interaction point, as shown in Fig.~\ref{fig:BTEV}. 
The second spectrometer of the BTeV is there to give them a factor of two in 
event yield, which is needed to partially compensate for the lower \bbbar\ 
cross section at the Tevatron.

The LHCb detector will have a planar geometry silicon vertex detector with 
r--$\phi$ strips. The Level--1 trigger of the LHCb will be high--\pt\ 
lepton(s) or hadron(s), while 
Level--2 incorporates vertex topology and track triggers~\cite{korolko}. 
In contrast, The BTeV collaboration plans to use silicon pixel detectors in 
their vertex detectors (also with a planar geometry) and relies
on a vertex--topology trigger in Level--1~\cite{simon}.

Both experiments will have RICH particle identification 
systems~\cite{brook,tomasz},
giving them a clear advantage over the general purpose detectors, ATLAS and CMS.

The LHCb collaboration submitted its Technical Proposal in March 1998, 
while BTeV was recently approved as an R\&D project to prepare a Technical 
Proposal. Since the proposed BTeV pixel and trigger designs are technically 
very challenging, an intensive R\&D program on these topics is the highest 
priority of the BTeV collaboration~\cite{penny}.

\section{R\&D projects}

The B experiments presently being constructed or undergoing upgrades and the 
future, next generation experiments all require major R\&D efforts,
for all sub--systems from vertex detectors to trackers, to trigger and 
data acquisition. There were many reports from these efforts at this 
conference, all reporting positive results. I will mention only a few of them, 
where the results are relevant for more than one experiment.
Readers should consult the individual contributions in these Proceedings
for more details.

\subsection{Micro--Strip--Gas--Chambers}
\label{sec:msgc}

During the last few years, 
Micro--Strip--Gas--Chambers~\cite{msgc} (MSGC) have become a promising tracking
device for high--rate, high--occupancy HEP experiments. They have been selected
as the "baseline solution" for CMS and HERA-B and are being considered by LHCb.
 
The initial R\&D on MSGC concentrated on aging problems which were solved by 
the application of a thin diamond--like coating on the glass 
substrate~\cite{diamond} and using clean materials for the assembly and clean 
gas systems.

The main problem of the last two years was the so called ``Sparking Problem'', 
in which strips are destroyed by gas discharges induced by heavily ionizing 
particles~\cite{viesbeck}, This problem was first reported by the HERA-B inner 
tracker collaboration at the Beauty96 conference at Rome~\cite{bernhard}. 
Since then, extensive R\&D on this subject was carried out by the CMS and 
HERA-B collaborations and two solutions were presented in this conference:

\begin{itemize}
\item CMS opted for ``advanced passivation''~\cite{passivation}, in which a 
thin layer of polyimid is applied to the cathode edges to prevent field 
extraction of electrons, and a lower voltage is used to achieve 
a gas gain of around 1000~\cite{manola}.

\item In HERA-B, higher gas gains are needed because the MSGC are used in
the trigger. The HERA-B solution, the GEM--MSGC~\cite{hottthesis}, is an 
alternative design based on the Gas Electron Multiplier (GEM)~\cite{sauli}. 
The GEM foil acts as an internal pre--amplifier and provides a second region of 
gas amplification. This allows the operation of GEM--MSGCs with gas gains of 
4000 or more, while both regions of gas amplification are still in a 
non--critical mode for gas discharge~\cite{hott}.
\end{itemize}

\subsection{Silicon microVertex detectors}
In order to minimize extrapolation distances and impact parameter errors, 
MicroVertex detectors of the forward geometry spectrometers are positioned 
inside the beam pipe close to circulating beams. The P238 
test--experiment~\cite{p238hard} demonstrated that such detectors could be 
reliably operated a few mm from the circulating beams. At high interaction
rates, radiation damage limits how close these detectors can approach the 
beams. However, in the future, advances in radiation hardness of silicon 
detectors or the use of new radiation--hard materials may improve the vertex 
resolution of the forward spectrometers by allowing them to run closer to the 
circulating beams. The vertex resolution of the forward detectors could also 
be improved by minimizing the multiple scattering of tracks.
There are two classes of dead materials which contribute to this problem:
\begin{itemize}
\item When detectors are placed inside the beam pipes, 
it is now a general practice to enclose them in thin ($\sim 100 \mu$) 
aluminium pockets to shield them from RF pick up from the circulating 
beams~\cite{p238hard,tasso,kirsten}. 
\item The large vertex vacuum vessels used in forward geometry detectors 
introduces abrupt changes to beam pipe cross sections which effect the 
impedance of the storage ring and introduce beam instabilities. In HERA-B, the 
thin metal ribbons or wires are placed around the beams inside the vertex 
vacuum vessels~\cite{tasso} to carry the image charge. 
\end{itemize}

Both of these additional materials increase multiple scattering in the vertex  
detectors. Additional R\&D, involving both accelerator and detector physicists,
is needed to determine if the aluminum RF shields around the silicon detectors 
can be eliminated or at least further reduced in thickness, to minimize their 
impact.

\subsection{Triggering trends in B experiments}

\BBbar\ events look very much like minimum bias events, apart from 
having detached secondary and possibly tertiary vertices and a somewhat higher 
transverse momentum of the B hadron decay products. Moreover, because the 
``interesting'' final states are a small fraction of the \BBbar\ events 
produced, hadron experiments are required to run with interaction rates of tens 
of megahertz.  The common strategy is to use  multi--level trigger schemes 
where at each stage more time, more data and more intelligence is introduced 
for the trigger calculations. The general trend in multi--level B triggers is 
to use high--\pt\ lepton and hadron triggers and store the data in pipeline 
buffers (on the detectors) during the Level--1 calculations. In the large 
general purpose experiments, triggers based on tracking and event topology are 
usually introduced in Level--2, while HERA-B employs tracking in Level--1 and
BTeV plans to trigger on event topology in Level--1. The Level--1 trigger 
calculations are performed with custom hardware; general purpose processors 
are employed as early as Level--2~\cite{mad}. The final trigger level performs 
full event construction. All experiments aim for archival rates of about 100~Hz.

Detailed algorithms for early trigger levels were developed and their expected 
performance were extensively studied with Monte Carlo event generators. The 
same is not true for the higher trigger levels where only some general ideas 
exist. More effort is needed in this area. 

There is a problem in hadron experiments which I call a ``dilemma of riches''.
Assuming all background from minimum bias could be eliminated at the trigger 
level, 100~Hz archival rates are barely sufficient for archiving all 
reconstructable B events. This will force collaborations to select (trigger on)
only certain B decay modes. Thus, the so--called "inclusive B experiments"
will be not quite inclusive.

\subsection{A "true" B--factory}

On the last day of the conference, Gerry Jackson talked on 
``A Dedicated Hadronic B Factory''~\cite{jackson}.
He pointed out that the Very Large Hadron Collider (VLHC) project, under study 
at Fermilab with a new superconducting transmission--line super--ferric magnet 
technology, will require a 3 TeV injector to allow the VLHC to reach 50 to 60 
TeV beam energies. The estimated cost of the 3~TeV machine is $\sim$370~M\$.

He then went on to say that this 3 TeV accelerator could be operated as a 
3 TeV pp collider (the magnets provide two dipole fields with opposite 
polarity). The machine parameters and the interaction region beam optics would 
be optimized to match the requirements of a dedicated B physics experiment. 

The 3 TeV machine would also be a highly valuable accelerator demonstration 
of the new magnet technology and the cost savings of the magnets and of modern 
tunneling techniques. This new machine would have a diameter of 34~km (somewhat
outside the FNAL site) and would be filled by the Main Injector.

Such a facility would be a true "B--factory" with 60~kHz \bbbar\ production 
rate at \fnallumi , compared to 3.6~Hz at the asymmetric $e^+e^-$ machines. 
Although the \bbbar\ cross section at $\sqrt s = 6$ TeV is about a factor of 
two lower than at LHC, optimizing the machine for B physics and being the 
"prime user" should more than compensate for this factor.
For example, when the LHC luminosity becomes \lhctoplumi , the beams in the 
LHCb interaction region have to be ``blown--up'' to decrease the LHCb 
luminosity to \lhcblumi . Having a very narrow bunch--crossing 
would allow B mesons to decay outside of the beam envelope which would 
be beneficial for triggering and eliminating combinatoric backgrounds. 

\section{Comparison of present and future B experiments}

Tables ~\ref{Comp1} and ~\ref{Comp2} compare the physics capabilities of 
the 1st and 2nd--generation experiments, respectively, in measuring the three 
angles of the Unitarity Triangle.  Even though the CLEO experiment could 
observe indirect CP--violation and obtain a constraint on the angle $\gamma$, 
it is not included in the tables, since it is not likely that they will measure
these angles directly. The numbers from the BABAR and CDF experiments are given
in Table~\ref{Comp1} as representatives of asymmetric $\rm e^+e^-$ and central 
hadron collider experiments, respectively; the physics reach of BABAR and 
BELLE, and of CDF and D\O\, are expected to be very similar~\cite{bondarwayne}. 
The first row shows the nominal start up year with full detector set up. The 
\bbbar\ production rates are calculated with the design luminosities given in 
the third row. In the remainder of the tables, values were calculated for one 
``Snowmass year'', $10^7$ sec. and using each experiment's ``design" 
luminosity\footnote{``Luminosity leveling'' planned for Run--II at Tevatron and 
also possible for the LHCb experiment will result in a better integrated 
Luminosity than assumed here.}.

The next two rows in the tables compare different experiments based on their 
expected errors in the so called, 
``$mixing$--$induced$ CP Asymmetry, $A^{mix}_{CP}$''. 
For the angle $\beta$, the \BdtoJK\ decay channel is used and therefore
the error on $A^{mix}_{CP}$  is equivalent to $\delta (sin(2\beta))$. 

Only the decay channel \Bdtopipi\ is considered for the angle, $\alpha$, and 
the experimental errors were scaled with a new assumed branching ratio of 
$BR(\Bdtopipi) = 7 \X 10^{-6}$. The low branching ratio effects both the 
number of signal events and the ratio of signal to background and 
therefore has a more detrimental effect on the central high--\pt\ experiments 
which have large backgrounds from the 2--body hadronic decays.
For this decay mode, the quoted errors on $A^{mix}_{CP}$ ignore  
penguin pollution and assume $A^{dir}_{CP} = 0$. With this assumption, 
$\delta (A^{mix}_{CP}) = \delta (sin(2\alpha))$.

No error on the angle $\gamma$ is given in the tables\footnote{The details 
of the angle $\gamma$ measurements and the expected errors are discussed in 
Ref.~\cite{neville} and \cite{penny}.} since the experiments could use 
more than one method to extract this angle and usually the error on the angle, 
$\gamma$, depends on other quantities such as \xs , the strong phase, etc.  

The next row shows the limits on \xs\ where the \Bs\ oscillations could be 
observed. And finally, the last row identifies the experiments which are 
capable 
of observing significant signals in the \Bstomumu\ decay mode with the expected 
Standard Model branching ratio of $BR(\Bstomumu) = 4 \X 10^{-9}$~\cite{buras}.

All 1st--generation experiments have sufficient sensitivity to observe CP 
violation in the \BdtoJK\ decay mode and measure $sin(2\beta)$.
The measurement of angle $\alpha$ is already difficult given the fact that the
\Bdtopipi\ decay mode suffers from penguin pollution. But even if penguin 
pollution could be kept under control, if the branching ratio of \Bdtopipi\ 
turns out to be small as assumed in these tables, the measurement of angle, 
$\alpha$, may be exceedingly difficult at the 1st--generation 
experiments. On the other hand, the BABAR and BELLE experiments could measure
angle $\alpha$ using other channels, such as \Bdtoropi . 
These possibilities have not been fully studied in the hadron experiments. 

Since the central high--\pt\ experiments have very limited or no particle
identification capability, the \Bdtopipi\ decay mode will suffer from large 
backgrounds from the \BdtoKpi , \BstoKpi\ and \BstoKK\ decay modes. 
In calculating the error on $A^{mix}_{CP}$, backgrounds from these 
2--body decay modes are assumed to have no CP asymmetry.
Although CDF and D\O\ may observe CP asymmetry in this topology,
due to the presence of these backgrounds it will probably be 
impossible for them to extract $\alpha$ from their measurements.

Although $e^+e^-$ experiments could observe ``direct CP violation'' and
set constraints on the angle $\gamma$, none of the 1st--generation experiments 
measure $\gamma$ using \Bs\ decay modes. 
Similarly, \xs\ limits at hadron machines are also limited by the detector 
resolution and event rates not much above the present LEP limit. 
$e^+e^-$ experiments could not contribute to \Bs\ physics, since they will
presumably be running at the $\Upsilon _{4S}$, at least in the first few years.

Table ~\ref{Comp2} shows a similar comparison for the 2nd--generation 
experiments using the latest information available at the time of the writing
this paper.  Since the Beauty97 conference, the BTeV collaboration introduced 
improvements to their silicon vertex detector geometry and studied the response
of their proposed topology trigger algorithm to multiple interactions; they 
reported that the algorithm is still functional at luminosities \fnallumi . 
Most of the improvement in \xs\ is due to use of a different decay mode than 
in the original study and an improved silicon vertex design~\cite{penny}.

The new LHCb numbers are from their Technical Proposal and include the
increase in event yields due to their new trigger implementation and the
the higher running luminosity~\cite{LHCB}. 

In the 2nd--generation experiments, due to the larger \bbbar\ production rate 
and/or more efficient triggers, the errors on $sin(2\beta )$ and
$sin(2\alpha )$ (only for the dedicated B experiments) decrease
to the level where stringent tests of the Standard Model could be made. 
The small differences between the ATLAS and CMS errors are 
due to the electron identification at lower momenta in ATLAS with their
Transition--Radiation--Detector tracker.
Both ATLAS and CMS suffer from the lack of particle identification for hadrons. 
Here again, asymmetry measurements by the general purpose high--\pt\ 
experiments using the \Bdtopipi\ decay mode will have 
limited use, due to the large 2--body B decay background. 
On the other hand, the dedicated B experiments, BTeV and LHCb, 
will not have this type of background. 
The only limitation on $\alpha$ from these experiments is the penguin 
pollution.

The BTeV and LHCb experiments, with their good particle identification 
capabilities, are the only experiments which have access to
the angle $\gamma$ using \Bs\ decay.

All 2nd--generation experiments could observe \Bs\ oscillations at much
larger values of \xs\ than could the 1st--generation experiments. 
They could also see a significant signal 
in \Bstomumu\ with the Standard Model branching ratio in a few years.

Of course the B physics programs of the 1st and 2nd--generation 
experiments are much richer and not limited to these decay modes. 
There are many other decay modes which could be used in the determination of
the angles or to look for New Physics effects.
Initially, measurements in different decay modes could be combined 
to decrease the errors on the CP--asymmetries. 
Eventually, these may provide additional tests of the Standard Model.

The Standard Model prediction of CP violation in \BstoJphi\ is very small, 
$A^{mix}_{CP} = 2\lambda^2\eta$. A measurement of the very small asymmetry
with good precision would therefore lead to a determination of $\eta$.
On the other hand, the observation of a large CP asymmetry in this channel 
could only result from New Physics.    

The very large heavy flavor data samples at hadron machines will be used to 
investigate the   spectroscopy of open flavor states such as \Bc\ and 
beauty--flavored baryons and hidden flavor \bbbar --onium states.
The ability of $e^+e^-$ experiments to construct final states containing 
$\pi^o$'s , i.e. \Bdtopzpz , is one of their strong points.

\section{Conclusions}

Despite the usual 
unforeseen problems in some experiments, all projects are moving ahead 
with full speed to observe CP violation in the decay, \BdtoJK , at the turn of
the century. However, the actual discovery time will depend on a few critical
issues. For example, when will the HERA-B tracking system be complete and when
will the $e^+e^-$ machines and the Tevatron collider reach their design 
luminosities ? 

If the branching ratio of \Bdtopipi\ turns out to be small as hinted by 
the recent CLEO data, the measurements of $sin(2\alpha$ ) might
take several years. In this case, experiments might observe ``Direct'' CP
violation in charged B decays before measuring $sin(2\alpha)$, thus
providing direct evidence for a non--zero value of the angle $\gamma$. 

It looks like we will have to wait for the 2nd--generation dedicated B 
experiments for the determination of all the angles. Even then there are 
discrete sign ambiguities to resolve. It is very important to measure these 
angles using a variety of decay modes. This will increase the chance of
observing New Physics effects, which otherwise could be missed. 

Since there are many theoretical assumptions in the methods used to extract 
the angles, the complete determination of the Unitarity Triangle and stringent
testing of the Standard Model will require independent determination of 
these parameters by a variety of methods using different decay modes. 
In this regard, rare kaon decay experiments on $ \rm K^o \ra \pi^o \nu\nu$ and
$\rm K^+ \ra \pi^+ \nu \nu$ will also contribute greatly.

There will be difficulties in the determination of the dilution factors, the
production asymmetries and the fake asymmetries possibly introduced by the
detector acceptance, triggers or reconstruction. All these problems can be 
overcome by studying large numbers of different final states. Although 
experiments always quote their CP--reach for $10^7$ secs, it certainly will 
take more than one year to understand the systematics of these measurements.

Finally, the present experimental programs have studied their B physics 
capabilities with only a few analysis methods using a handful of decay modes. 
As data from the new experiments become available, new analysis methods will
be cooked up and ways found to reconstruct decay modes which were thought to 
be very hard or impossible. There is no better creative stimulation than 
having real data to play with.

When all the work is done, there are the following possible outcomes:
\begin{itemize}

\item The CKM phase is zero and the Unitarity Triangle is flat. 
This is not very likely; although the present data does not rule out $\eta = 0$,
it is at the edge of the allowed region.

\item All measurements agree with the Standard Model description of CP 
violation and the angles, $\alpha , \beta$ and $\gamma$ sum to $\pi$.
Thus, there is no sign of New Physics. This would be very ``boring".

\item The Unitarity Triangle is not unitary, or it can not explain all the 
observed CP violation effects. Hence, we have signs of New Physics.
This would be very exciting.

\item However, the {\it most} exciting thing would be to 
find something totally unexpected which challenges all known theories.

\end{itemize}
During the last decade, discussing (and calculating) the merits of different 
ways to do B physics in our cyber experiments dominated our activities.
\begin{enumerate}

\item Asymmetric $e^+e^-$ vs. hadron machines ?

\item Fixed--target (Gas jet, wire targets, extracted beams) vs. collider ?

\item Forward vs. central ?

\end{enumerate}

One of the aims of this series of Beauty conferences was to find the answers to
these questions. The other was to create a forum to discuss developments in
the critical detector, trigger and data acquisition R\&D issues. Items 2 and 3 
have been resolved in favor of forward collider experiments. 
The community is slowly realizing that hadronic B--experiments will
play a more and more dominant role in the B physics area.
Soon real data will be available and the real fun will begin.

The next decade will be even more exciting if, as has happened before, data 
leads the theory, rather than following and confirming theoretical predictions.

\section{Acknowledgements}
I would like to congratulate all the speakers for their excellent talks and 
thank them for providing me with their results and transparencies. 
I would also like to thank all the participants for very useful and 
stimulating discussions over the coffee breaks, during the walks to/from hotels
and meals. Some of these lively discussions even continued at the
conference dinner.

I think I reflect the sentiments of all the participants when I thank
our hosts, Peter Schlein and Jim Kolonko, for organizing a very 
smooth and pleasant workshop and making this conference a very enjoyable
experience for all of us. 

\clearpage


\clearpage

\section*{TABLES}

\begin{table}[h]
\centering
\caption[]{List of B physics experiments planned or under construction. 
Next generation hadron experiment are preparing technical 
design reports of their detector sub-systems (TDR)
or technical proposals (TP).}         
\label{bprogram}
\vspace{3.0mm}
\begin{tabular}{||l|l|l|r|l||}
\hline
Exp. & Lab. & Accelator & date & Status \\
\hline
CLEO III & Cornel & Sysm. $e^+e^-$ & 1999 & Upgrade \\
BABAR    & SLAC   & Asym. $e^+e^-$ & 1999 & Construction \\
BELLE    & KEK    & Asym. $e^+e^-$ & 1999 & Construction \\
\hline
HERA-B   & HERA & wire target & 1999 & Construction \\
CDF      & FNAL  & Tevatron &2000 & Upgrade \\
D\O      & FNAL  & Tevatron & 2000 & Upgrade \\
\hline
BTeV     & FNAL  & Tevatron & $\sim$2003 & Approved R\&D for TP \\
ATLAS    & CERN  & LHC & 2005 & TDRs are partially approved \\
CMS      & CERN  & LHC & 2005 & TDRs are partially approved\\
LHCb     & CERN  & LHC & 2005 & Submitted TP, March 1998 \\
\hline
\end{tabular}
\end{table}

\begin{table}[h]
\centering
\caption[]{Physics comparison of the 1st--generation B experiments for 
$10^7$ s, as discussed in the text.}         
\label{Comp1}
\vspace{3.0mm}
\begin{tabular}{|l||c|c|c|}
\hline
                    & HERA-B  & BABAR          & CDF   \\
                    &         & BELLE          & D\O\  \\
\hline
Year                & 1999    & 1999           & 2000  \\
\bbbar\ rate        & 40 Hz   & 3.6 Hz         & 2 kHz \\
\lum               &         & $3 \x 10^{33}$ & $2 \x 10^{32}$ \\
\hline
Ang($\beta$)        & 0.13    & 0.16           & 0.08  \\
Ang($\alpha$)       & 0.24    & 0.26           & 0.26  \\
Ang($\gamma$)       & No      & No             & No    \\
\xs                 & 17      & No             & 20    \\
FCNC (\Bstomumu )   & No      & No             & No    \\
\hline
\end{tabular}
\end{table}

\begin{table}[h]
\centering
\caption[]{Physics comparison of the 2nd--generation B experiments for 
$10^7$ s, as discussed in the text. All numbers given in this Table are
the latest values received from the collaborations~\cite{butler,LHCB,LHCC}
(March 1998).}         
\label{Comp2}
\vspace{3.0mm}
\begin{tabular}{|l||c|c|c|c|}
\hline
                  & B-TeV                 & LHCb             & ATLAS
                        & CMS \\
\hline
Year              & $\sim 2003$               & 2005             & 2005 
                        & 2005 \\
\bbbar\ rate      & 20 kHz         & 100 kHz           & 500 kHz 
                        & 500 kHz \\
\lum              & 2$ \x 10^{32}$ & $2 \x 10^{32}$ & $1 \x 10^{33}$ 
                        & $1 \x 10^{33}$ \\
\hline
Ang($\beta$)      &  0.016          & 0.011             & 0.017
                        & 0.021 \\
Ang($\alpha$)     & 0.04          & 0.05             & 0.18
                        & 0.17 \\
Ang($\gamma$)     & Yes                   & Yes              & No
                        & No   \\
\xs               & 70              & 75              & 38
                        & 38   \\
FCNC (\Bstomumu ) & Yes                   & Yes              & Yes
                        & Yes  \\
\hline
\end{tabular}
\end{table}

\clearpage

\subsection*{FIGURE CAPTIONS}

\end{document}